\documentclass[preprint,showpacs,floats,letterpaper,floatfix,,groupedaddress,eqsecnum]{revtex4}

\usepackage{subfigure}

\usepackage{amssymb,amsmath}
\usepackage[dvips]{graphicx}
\usepackage{csquotes}

\usepackage{dcolumn,epsfig}
\usepackage[left]{lineno}

\begin{document}

\title{Stochastic resonance in a sinusoidal potential system: An analog 
simulation experiment}

\author{Ivan Skhem Sawkmie$^1$, and Mangal C. Mahato$^1$$^,$}
\email{mangal@nehu.ac.in}
\affiliation{$^1$Department of Physics, North-Eastern Hill University,
Shillong-793022, India}

\begin{abstract}
Recently, stochastic resonance was obtained numerically in an underdamped periodic potential system driven by a periodic force and a Gaussian white noise. In that numerical work, the occurrence
of stochastic resonance was explained in terms of the existence of two dynamical states having different amplitude and phase lag. At zero temperature these two initial condition dependent dynamical states are stable. However, at elevated temperatures, these two states make transitions from one to the other at a mean rate.
In the present work, we setup an analog simulation experiment to show the existence of the two dynamical states in a sinusoidal potential system as well as to verify 
the occurrence of stochastic resonance in the same system. The experimental procedure includes setting the initial conditions for the experiment.
\end{abstract}

\vspace{0.5cm}
\date{\today}

\pacs{07.50.Ek, 05.40.jc, 05.60.Cd, 05.10.Gg, 05.45.-a}
\maketitle

\section{Introduction}

Stochastic resonance (SR) is a phenomenon wherein the response of a nonlinear system to a subthreshold periodic input signal can be optimized by appropriately tuning the noise
intensity\cite{Benzi}. The output characteristics of the system, such as the 
signal-to-noise ratio (SNR), have a well-marked maximum at the optimal noise level. The phenomenon of 
SR is usually found to occur theoretically as well as experimentally in 
bi-stable (or double-well) potential systems\cite{Gammaitoni}; some of the early examples 
include the model Landau potential (theory), the Schmitt trigger circuit, and
two-mode ring lasers (experiment)\cite{McNamara,Fauve,Wiesenfeld}. 
However, later, SR was found to occur in monostable systems as well. The 
first observation of SR in monostable systems was in 
a tilted single well Duffing oscillator driven by a weak periodic force and a 
Gaussian white noise\cite{Dykman,Luchinsky,Stocks,Stein}.

In 1983, SR was studied experimentally in the Schmitt
trigger system where the SNR was first used to describe the
phenomenon\cite{Fauve}. The Schmitt trigger circuit is particularly 
interesting because it is explained in terms of a discrete 
two-state system with hysteresis. It was shown that the SNR, at the
output of the Schmitt trigger subjected to a weak periodic signal in presence of noise, 
increases with increasing noise intensity,
passes through a maximum and then decreases. In the experiment, the output voltage switched randomly from one state to the other periodically at the external drive frequency when the amplitude of the noise strength was increased. That is, the power spectrum became maximum around the forcing frequency at an optimal noise level showing SR.

Recently, it was numerically found that SR occurs in periodically driven underdamped periodic potential systems\cite{Saikia,wanda2012}. In these numerical works, instead of the (output) signal-to-noise ratio, an equivalent measure of the input energy, absorbed per period of the drive, was used\cite{Mahato,Sekimoto,Evstigneev}. It was found that for a given sinusoidal drive one obtains two kinds of particle trajectories, one with a small amplitude (SA) having a small phase lag behind the drive and the other with a large amplitude (LA) and a large phase lag. At $T=0$, it is the initial condition that determines the state of a trajectory, that is, for a given drive, some initial conditions give SA whereas the others give LA. The trajectories SA and LA have the status of dynamical states\cite{Saikia}. The occurrence of SR was shown to be related to the existence of the two dynamical states of (output signal) trajectories and transition between them as the noise strength (temperature $T$) is increased.  The two dynamical states were found to coexist only in a restricted domain of parameter space of the friction coefficient $\gamma$ of the medium and the angular frequency $\omega$ of the (input) periodic sinusoidal drive (signal). Since the output signal is periodic with the same frequency $\omega$ but lag behind the input signal by a phase, the system shows hysteretic behaviour; a plot between these two shows a hysteresis loop. The area of the hysteresis loop so obtained represents the (input) energy absorbed by the system from the external drive (and also lost to the environment) per period of the drive field. The variation of hysteresis loop area as a function of noise strength indicates the occurrence of SR in sinusoidal potentials. However, so far no experiment has been conducted to show the occurrence of SR in sinusoidal potentials. In this work, we setup an analog simulation experiment to verify the occurrence of SR in sinusoidal potentials. Our experiment is similar to the analog simulation work done earlier to study stochastic nonlinear dynamics\cite{McClintock,Luchinsky2}. 

 A driven damped simple pendulum provides a useful but exact mechanical analogy to our analog simulation experiment. A periodically driven simple pendulum has the same equation of motion as the motion of a
particle in a sinusoidal potential. As the amplitude of a simple pendulum is not restricted to small values, its motion differs from that of a harmonic oscillator. In the same way, our analog electronic circuit experiment differs from an ac driven LCR circuit experiment. In Fig. 1, the periodic drive is applied at the point A. As a part of the expected scenario, depending upon the initial position of release of the pendulum, the pendulum provides only two distinct solutions: one in which the pendulum oscillates with a higher amplitude and having a larger phase lag (LA state) with respect to the periodic drive, and the other with a smaller amplitude and having a smaller phase lag (SA state). We verify these results using our analog simulation experiment.

\begin{figure}[htp]
\centering
\includegraphics[width=4cm,height=4cm]{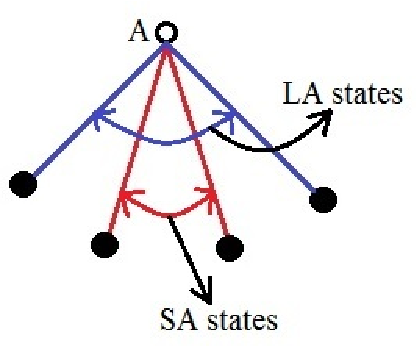}
\caption{The schematic simple pendulum}
\label{fig1}
\end{figure}

In the next section, we give a gist of the numerical work where SR occurs in a sinusoidal potential system. In Section III, we give details of the experimantal setup and also describe a procedure to set the initial conditions for the analog simulation. In section IV, we explain the results from our experimental work and in Section V, we discuss and draw conclusion.

\maketitle
\section{SR in sinusoidal potential systems (Numerical Work)}

In the numerical work\cite{Saikia}, a particle of mass $m$ moving in a periodic potential 
$V(x)=-V_0\cos(kx)$ in a medium with friction coefficient $\gamma$ and 
subjected to an external periodic forcing $F(t)=F_0\sin(\omega t)$ is 
considered and described by the Langevin equation
\begin{equation}
m\frac{d^{2}x}{dt^{2}}=-\gamma\frac{dx}{dt}-\frac{\partial V(x)}{\partial x}+F(t)+\sqrt{2\gamma T} \xi(t)
\end{equation}

The temperature $T$ is in units of the Boltzmann constant $k_B$. The random fluctuations $\xi(t)$ satisfy the statistics $<\xi(t)>=0$ and $<\xi(t)\xi(t')>=\delta (t-t')$. Eq. (2.1) is written in dimensionless units by setting $m=1$, $V_0=1$ and $k=1$ as,
\begin{equation}
\frac{d^{2}x}{dt^{2}}=-\gamma\frac{dx}{dt}-\sin(x)+F(t)+\sqrt{2\gamma T} \xi(t)
\end{equation}

The solution is obtained in a limited range in the $(\gamma,\omega)$ space with 
$F_0=0.2$. Note that due to the external force, $F(t)=F_0\sin(\omega t)$, the 
potential, $-\cos(x)$, gets tilted by an effective slope of $F_0\sin(\omega t)$. 
In the extreme cases, the effective potential becomes $V_{eff}=-\cos(x)\pm F_0x$, so that the potential 
barrier disappears, momentarily or for a longer period, only for $F_0\ge1$. 
$F_0=0.2$, thus corresponds to the potential barrier remaining finite 
throughout the period of $F(t)$.

Interesting numerical results were obtained in the range, $\omega\approx1$ 
and $0.07\le\gamma\le0.16$, of parameter space $(\gamma,\omega)$ or 
$(\gamma,\tau)$, with $\tau=\frac{2\pi}{\omega}$. Since the potential is 
sinusoidal, the choice of the initial positions $-\pi\le x(0)<\pi$, suffices 
to obtain all possible trajectories, $x(t)$, $t> 0$ with the initial 
velocity set equal to zero. It was found that the trajectories of the 
particles have two distinct solutions, for the same drive $F(t)$, which assume the status of dynamical 
states. These two dynamical states are distinguished by the lag in phase 
$\phi$ of the response with respect to the externally applied sinusoidal 
field. In one state (in-phase, SA) $\phi$ is small and in the other (out-of-phase, LA) 
$\phi$ is large. Interestingly, the out-of-phase dynamical state has a 
large amplitude compared to that in the in-phase state, so that the 
particle explores almost the entire span of the sinusoidal well. The 
probability of getting either of the two states depends on the initial 
condition, especially at low temperatures. As the temperature is increased 
further, at some moderate temperature, the relative population of the two 
states become almost equal and the input energy peaks, thus, obtaining SR in the 
sinusoidal potential \cite{Saikia,wanda2012,wanda2013,wanda2015,donrich2016}. The input energy expended per period of the 
external field on the system acts as a good quantifier of 
SR\cite{Iwai,Evstigneev}. In the following, we present our experimental work to verify the above 
numerical results.

\maketitle
\section{Experimental Setup}

\subsection{Circuit Model}
In order to simulate the equation (2.1) or (2.2), we set up an electronic 
circuit shown in Fig. 2. The system is driven periodically by an external 
periodic input current $I_{inp}(t)=I_0\sin(\omega t)$. The circuit simulates 
(at A) the Langevin equation:
\begin{equation}
R_2C_1C_2\frac{d^{2}V_{out}}{dt^{2}}=-\left\{ \frac{R_2C_1}{R_B}+C_3+\frac{R_2C_2}{R_A} \right\} \frac{dV_{out}}{dt}-\frac{U_0}{V_0}
\sin\left(\frac{V_{out}}{V_0}\right)+\frac{V_{in}(t)}{R_1}+\xi(t)
\end{equation}

\begin{figure}[htp]
\centering
\includegraphics[width=14cm,height=6cm]{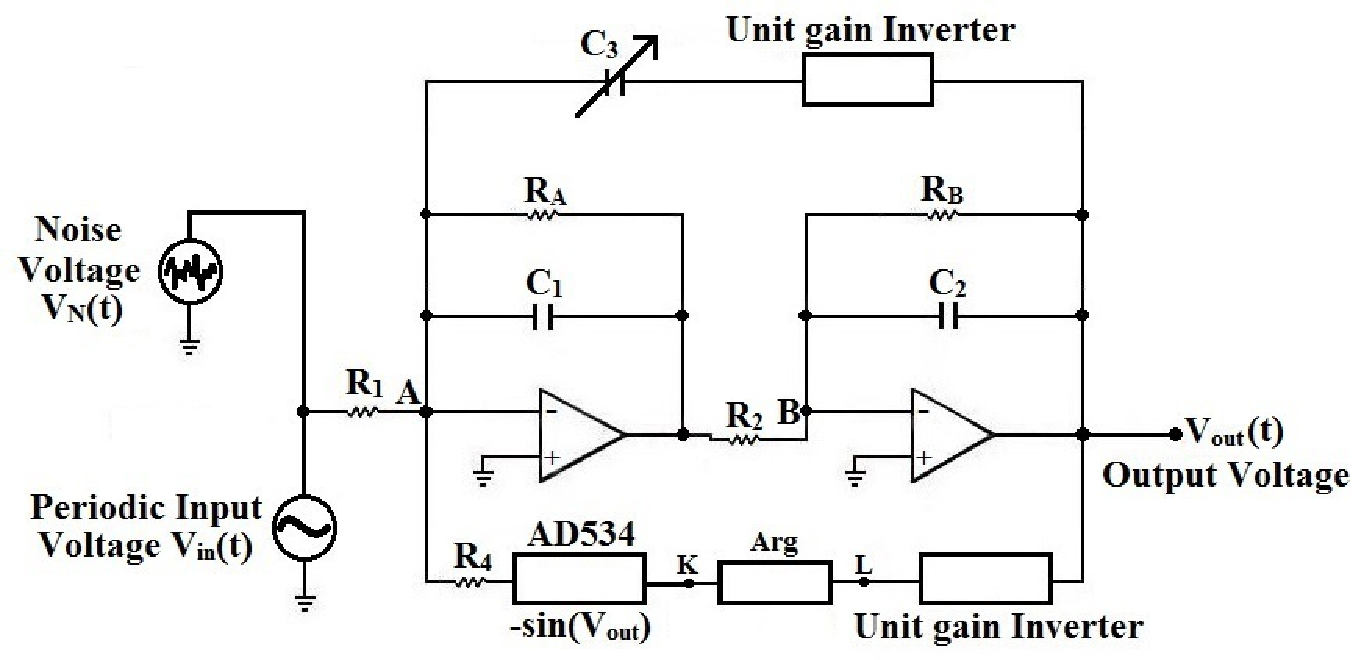}
\caption{Block diagram to simulate Eq. (3.1). Here the parameters $R_1=5.1K\Omega$ when $V_{in}\approx$ 202mVpp, $R_2=5.1K\Omega$, $R_4=10.0K\Omega$, $R_A=1M\Omega$, $R_B=470K\Omega$, $C_1=1.0nF$ and $C_2=10.0nF$ are fixed parameters whereas $C_3$ is a variable parameter (for example, $C_3$ is set equal to $212.47pF$ for $\gamma =0.1181$). The block Arg is elaborated in Fig. 3.}
\label{fig2}
\end{figure}

Here we have taken $V_0=1$volt, $U_0=\frac{V_{0}^2}{R_4}$volt$^2$/ohm and 
$\xi(t)=\frac{\sqrt{4TR\Delta f}}{R_1}$ ampere, $\Delta f$ is the bandwidth of the noise signal and $\xi(t)$ is the (Gaussian) current fluctuation. The rms voltage of the noise signal is $V_{rms}=\sqrt {4k_BTR\Delta f}$volt. The random fluctuations $\xi(t)$ or $\zeta(t)=\sqrt{\Delta f}$ sec$^{-1/2}$, satisfy the statistics 
$<\zeta(t)>=0$ and $<\zeta(t)\zeta(t')>=\delta (t-t')$.
Here $T$ is the temperature (noise strength) measured in units of the Boltzmann constant $k_B$. Eq. (3.1) is written in dimensionless units \cite{desloge} by setting $'m'=R_2C_1C_2=1$, $U_0=1$ and $'k'=\frac{1}{V_0}=1$. The Langevin equation with reduced variables denoted again by the same symbols, corresponding to Eq. (3.1) is written as
\begin{equation}
\frac{d^{2}V_{out}}{dt^{2}}=-\left\{\frac{R_2C_1}{R_B}+C_3+\frac{R_2C_2}{R_A}\right\}\frac{dV_{out}}{dt}-
\sin(V_{out})+\frac{V_{in}(t)}{R_1}+\frac{\sqrt{4TR}}{R_1}\zeta(t)
\end{equation}

Comparing Eqs. (3.2) and (2.2), we see that these two equations are similar with damping coefficient $\gamma =\{\frac{R_2C_1}{R_B}+C_3+\frac {R_2C_2}{R_A}\}(R_4/m)^{0.5}$. 
Here $V_{in}(t)=V_{in}^0\sin(\omega t)$ where $V_{in}^0$ is the amplitude of the input (signal) voltage, 
$\omega=2\pi f$ ($f$ is the frequency of the periodic input current). The temperature $T$ in dimensionless units is calculated using the equation,

\begin{equation}
T=\frac{V_{rms}^{2}R_4^\frac{1}{2}}{4R\Delta f(R_2C_1C_2)^\frac{1}{2}V_0^2}
\end{equation}

Here, 'R' value is calculated by equating the coefficient of $\xi(t)$ from Eq. (2.2) with the coefficient of $\zeta(t)$ from Eq. (3.2).

The sinusoidal input voltage and the Gaussian noise that we have used in our experiment are taken from the Agilent 33500B series waveform generator. The oscilloscope that we have used in our experiment is the InfiniiVision, MSO-X 3014A from Agilent Technologies.

\subsection{Sine converter}

In the circuit model shown in Fig. 2, we need to take the sine of the output signal. For this purpose, we have used the IC AD534. We use the circuit design for the AD534 as specified from the datasheet. However, experimentally we find that this particular IC works best when the input signal is in the range from 0V to 1.165V out of the maximum 10V specified in the datasheet. Since this IC can perform only $\approx 12\%$ of the maximum input voltage, we modify the parameters related to AD534 so that the input signal to the sine converter (say $V_z$) can go beyond the available $\approx 12\%$ of the maximum input voltage. Here, we use trial and error method and arrive at a conclusion that, for different combinations of the parameters related to AD534, the output from the sine converter (say $V_{sine}$) should be of the form
\begin{equation}
V_{sine}=4.6\sin(\frac{\pi}{2} \times \frac{V_z}{5})
\end{equation}
where $V_z$ goes from 0V to $\pm$5V.
From Equation (5), we found that even if $V_z$ goes to $\pm$5V, the AD534 still gives approximately the required output signal $V_{sine}$. Therefore, we have a sine converter where the argument $\theta = ({\pi}/{2}) \times ({V_z/}{5})$  is in the range [$\frac{-\pi}{2}, \frac{\pi}{2}$]. In Fig. 2, the scale factor 4.6 from Eq. (3.4) has been compensated so that the output from the sine converter becomes $-\sin(V_{out})$.
  
\begin{figure}[htp]
\centering
\includegraphics[width=9cm,height=6cm]{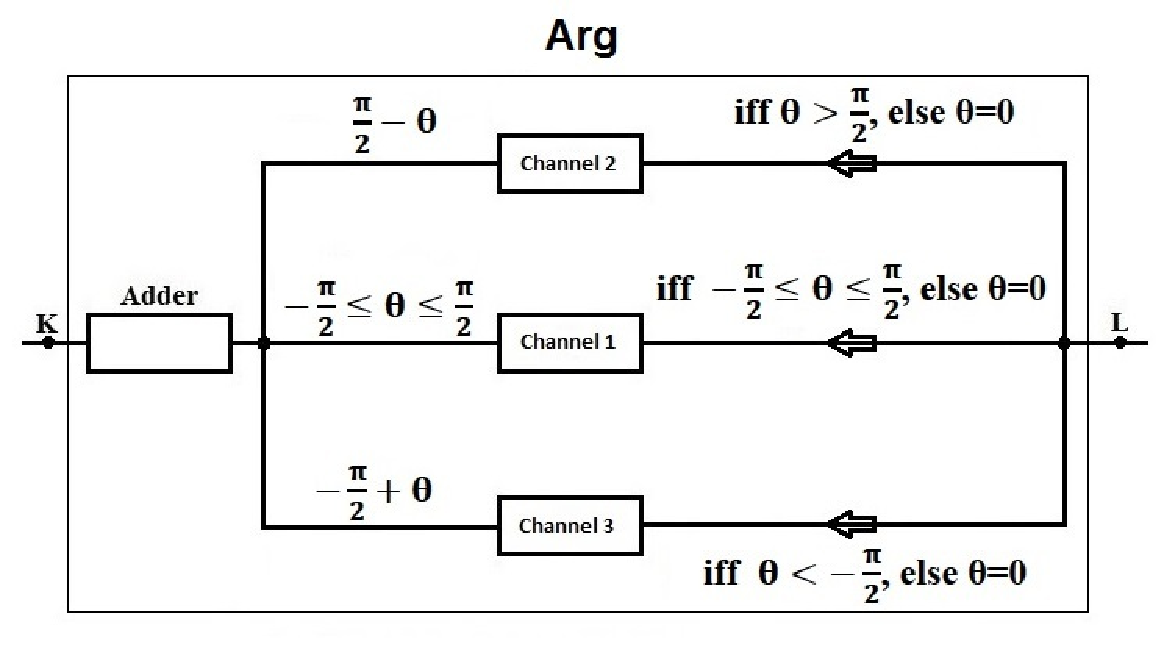}
\caption{Argument of sine to get the full potential}
\label{fig3}
\end{figure}

However, in order to get the full potential, $-\cos(\theta)$, with  $-\pi\leq\theta < \pi$, we have built an extra circuit (Fig. 3) consisting of three seperate channels:
\begin{enumerate}
  \item Channel 1: If $( -\pi/2 <\theta < \pi/2)$, then this channel allows the signal to go directly to the sine converter. If however, $( -\pi/2 >\theta > \pi/2)$, then the output signal from this channel is set equal to zero.
  \item Channel 2: If $( \theta > \pi/2)$, then we invert only the signal where $( \theta > \pi/2)$ with a reference value of $+\pi/2$. To achieve this result, we convert all the signals with $\theta>\pi/2$, say $\theta=\pi/2 + \theta'=V_+^1$ (where $\theta'=$+ve number), to $\theta=\pi/2 - \theta'$. In the circuit, for signals $>\pi /2=V_+^1$, we convert the signals to $(\pi -V_+^1)$ ($=V_+$, say) and we take $\sin (V_+)$ and this satisfies the condition $\sin \theta = \sin (\pi -\theta) $. If however, $(\theta < \pi/2)$, then the output signal from this channel is set equal to zero.
  \item Channel 3: If $( \theta < -\pi/2)$, we perform a similar procedure as for the second channel by inverting only the signal where $( \theta < -\pi/2)$ with a reference value of $-\pi/2$. To achieve this result, we convert all the signals with $\theta<-\pi/2$, say $\theta=-\pi/2 - \theta'=V_-^1$ (where $\theta'=$+ve number), to $\theta=-\pi/2 + \theta'$. In the circuit, for signals $<(-\pi /2)=V_-^1$, we convert the signals to $(-\pi -V_-^1)$ ($=V_-$, say) and we take $\sin (V_-)$ and this satisfies the condition $\sin (-\theta) = \sin (-(\pi -\theta)) $. If however, $(\theta > -\pi/2)$, then the output signal from this channel is set equal to zero.

\end{enumerate}
  
Using the above three complementary channels, we obtain a sine converter where its output is approximately in the full range of $-\pi<\theta<\pi$.

\subsection{How to set the initial condition?}

The nonlinear (feedback) oscillator simulates a second order ordinary 
differential equation giving the solution in the form of an output voltage,
$V_{out}(t)$, in response to an input current, $I_{inp}(t)$. We solve the 
equation as an initial value problem. Naturally, we require to set two initial 
values at $t=0$. In our case, we obtain as solution $V_{out}(t\neq 0)$ given 
$V_{out}(t=0)$. We leave the other initial condition $\frac{d}{dt}V_{out}(t=0)$ arbitrary. In
the following we describe how we set the initial condition $V_{out}(t=0)$. 

Our electronic circuit is a weakly damped periodically forced oscillator.
Therefore, once the input signal $I_{inp}(t)$ is switched off, it will oscillate 
on its own with a gradually diminishing amplitude at its characteristic 
frequency \cite{johannessen}. Note that our drive frequencies are not far away from the 
characteristic natural frequency of the oscillator. We seek to choose a 
point on the freely oscillating first cycle as the initial condition 
$V_{out}(t=0)$. As per the design of the problem, we have a potential function 
$U(x)=-\cos(x)$, $x\equiv V_{out}$. Therefore, ideally, $V_{out}(0)$ must be 
obtained from a uniform distribution of values in the range, 
$-\pi\leq V_{out}(0)<\pi$. However, given the small amplitudes of $I_{inp}$, 
the amplitudes of the output signal $V_{out}$ at $T=0$ are much smaller than 
$\pi$. We could obtain the amplitude of the first cycle of the freely 
oscillating $V_{out}$, at most, roughly in the range of 
$-0.66\pi\leq V_{out}(0)<0.66\pi$ only. This is a major limitation of our
procedure.

Note that for periodic $I_{inp}(t)$ of frequency $f$ one is expected
to obtain a periodic $V_{out}(t)$ of the same frequency given the other
parameters suitably fixed. In the present case the 'other parameters' include 
$\gamma$ and the noise strength $T$. $I_{inp}(t)$, in turn, is obtained from  
$V_{in}(t)$ derived from a function generator (FG) in the form of sample points 
at a desired sampling rate (SRT). In the experiment, the solutions are obtained 
for various fixed values of $f=f_3=\frac{2\pi}{\tau}$, where $\tau$ is the period of the input signal $I_{inp}(t)$.

The whole procedure of setting the initial conditions consists of three parts
of which the initial two parts set the initial condition for the actual experiment, namely, the third part.
However, certain operational parameters in the first part are adjusted 
depending on the actual experimental conditions of the third part. For example, 
the sampling rate is kept fixed in the first two parts as in the third part. 
In the third part of the experiment, the total number of cycles ($TNC3$) of 
$V_{in}(t)$ and the total number of sample points ($TSP3$) in those number of 
cycles are kept fixed for all sets of experiment. That is, the number of sample
points per cycle is kept fixed. In other words, the sampling rate 
$SRT=\frac{f\times TSP3}{TNC3}$ is kept fixed for a given frequency $f$ of
$I_{inp}$. For a given amplitude of $I_{inp}$, the amplitude of the output 
signal $V_{out}$ depends on the frequency $f$ of the input signal. With these
information in mind we set about fixing the initial condition $V_{out}(t=0)$, as follows, in two steps. The actual experiment of obtaining $V_{out}(t)$, in response to an input signal
$I_{inp}(t)$ of frequency $f=f_3$, is carried out for which fixing the initial
condition $V_{out}$ is done at the beginning of the third part of the whole 
precedure.

The first step consists of trial runs to obtain $V_{out}(t)$ for $TNC3=18000$ 
cycles of the input signal consisting of $TSP3=1800\times 10^3$ sample points 
(or 1800 KSa) for various frequencies $f$. The sampling rate $SRT3=\frac{f\times TSP3}{TNC3}$ for frequency $f$ is fixed through the function generator (FG). For these
trial 
runs we use 1000 cycles of the $I_{inp}(t)$ of frequency $f$, as the 
first part of the procedure, and then set $I_{inp}=0$ for the second part. We, 
then, let the oscillator have free oscillations for a sufficiently long time 
so that oscillation amplitude of $V_{out}$ becomes zero before the beginning 
of the third part. Then, we switch on $I_{inp}(t)$ so that the third and 
final part of the trial run begins with $V_{out}(t=0)=0$ and continues for the 
entire $TNC3$ cycles of $I_{inp}(t)$. One can expect $V_{out}$ to be either in 
the large amplitude (LA) state or in the small amplitude (SA) state depending on the frequency $f$ of $I_{inp}(t)$. We thus
have a list of frequencies and corresponding amplitudes. From this list we can find
out the largest amplitude of $V_{out}(t)$ and we select 
a frequency $f=f_1$ for which the amplitude of $V_{out}$ is a little smaller 
than the maximum. We fix this selected $f=f_1$ as the frequency of $I_{inp}(t)$
for the first part of the procedure in the actual experiment. This completes the first step of our
procedure. 

In the final step which is our actual experiment, we select a particular $\tau $ value having a particular frequency ($f_3$) for the third part. This $\tau $ value is then calculated in terms of sampling rate (SRT3) and we keep SRT3 same as for $f=f_3$ in all parts, first, 
second, and the final third. For the first part ($f=f_1$, need not be equal to $f_3$), we take 100 KSa 
points so that it consists of $NC1=\frac{f_1\times 100\times 10^3}{SRT3}$ 
cycles. We take NC1, the nearest integral value of the calculated one, as the 
number of cycles of $I_{inp}$ of frequency $f_1$ in the first part of the 
procedure. After NC1 cycles of $I_{inp}$ is completed, the $I_{inp}(t)$ is
switched off so that $V_{out}$ begins free oscillation at the natural frequency
of oscillation of the circuit. The second part of the precedures begins from
that instant. We exploit the first cycle of the free oscillation to tap 
$V_{out}(t=0)$ for the third (final) part of the experiment. Note that, since
the frequencies $f$ of $I_{inp}$ are not far from the natural free oscillation
frequency, there are roughly 100 sample points per cycle of oscillation.
If, for example, we wish to use 50 different initial conditions $V_{out}(t=0)$
in our actual experiment, we choose the $2^{nd},~4^{th},~6^{th},
~\cdots$ sample points of the second part as the starting point for the third
part of the procedure by switching on the $I_{inp}(t)$ (now $f=f_3$) from 
$t=0$.

In practice, we control our BenchVue-enabled waveform generator (FG) and design a 
customized waveform (arbitrary signal) from our PC using the BenchLink 
waveform builder software. This arbitrary signal (of desired amplitude, for
example, 2V peak to peak) is then fed to the waveform 
generator either directly from the PC or through a pen drive. As per the 
procedure described earlier, the arbitrary signal consists of three parts. 
While designing the arbitrary signal from our PC, we have fixed the sampling 
rate (SRT) to 20KSa/s for all the three parts but this sampling rate can be 
varied depending on the frequency of the input signal. However, the frequency 
can be controlled directly from the waveform generator through the sampling rate. The practical 
implementaion of the procedure is illustrated in the following.

Let the total number of sample points in the first part of the procedure 
TNSP1 be 100KSa so that the duration of this part (TNSP1/SRT) equals 5s.
Moreover, if the first part contains TNC1=1000 cycles, then the frequency of 
the input signal = 200 Hz and the number of sample points per cycle equals
100 Sa/cycle. In order to have the same frequency in the third part also, we
take TNSP3=1800KSa and TNC3=18000. Now, if, for example, we wish, instead, to have a 
frequency of 250 Hz of input signal we change the SRT to 25 KSa/s with the 
help of the FG keeping all other numbers same as earlier. Note that in the 
process the duration (TNSP1/SRT) of the first part changes from 5s to 4s and,
similarly, the duration (TNSP3/SRT) of the third part changes from 90s to 
72s. For the trial runs, we change the frequencies $f$ of the input signal in 
a similar fashion with the help of the FG and measure the corresponding 
amplitudes $V_{out}(t)$ of the output signal at the end of the third part. In these trial runs, the
initial condition $V_{out}(0)=0$ is ensured in the beginning of the third part 
by letting the free oscillation for a sufficiently long time in the 
second part. We choose $f=f_1$ for which the amplitude was a little smaller
than the maximum of all $V_{out}(t)$ of the trial runs. We then proceed to
the actual experiment as the last step.  

For the actual experiment with signal frequency $f_3$ in the third part, 
we set the sampling rate
SRT=TNSP3/(TNC3/$f_3$)=$f_3\times 100$ Sa/s. Note that we have kept TNSP3 and
TNC3 same (1800 KSa and 18000, respectively) as earlier. This SRT is same for 
all the three parts. For the first part we keep TNSP1 same (100 KSa) as earlier
but now TNC1 is calculated taking the signal frequency of the first part to be 
the chosen frequency $f_1$ obtained from the trial runs. Thus, TNC1=
$\frac{f_1}{f_3}\times 1000$. We, however, take the closest integer as the 
actual TNC1. Note that the duration of the first part is $\frac{1000}{f_3}$. 
The rest of the procedure of obtaining the $V_{out}(0)$ follows as 
described earlier.

\subsection{Signal Acquisition and Analysis}

Analysis of the behavior of the circuit model usually involves two main stages: digitization of the analog signal (i.e., $V_{out}(t)$ and $V_{in}(t)$) and then processing of the resultant digital time series to extract the required particular information. $V_{out}(t)$ and $V_{in}(t)$ are taken directly from the MSO and saved in the pendrive in ".csv" format where each file consists of 5,00,000 data points. From the FG, we have used a burst mode where the FG allows one run (which consists of the first, second and third part of an arbitrary signal), stops for few millisecond before it allows again for another run. This process continues as long as the input channel and the burst mode are switched on. From the oscilloscope (MSO), we save the number of cycles much larger than the required number of cycles for analysis. That is, in a given single run in the MSO, we save $R_{123}$ cycles consisting of the cycles in the first, the second and the third part of our initial condition setting procedure (of the arbitrary signal) and also $R_{1-}$ cycles preceding $R_{123}$ cycles and $R_{3+}$ cycles succeeding the $R_{123}$ cycles. These 5,00,000 data points, thus, consist of the $R_{123}$ cycles plus the $R_{1-}$ and $R_{3+}$ cycles.

Processing of the signals is done in the PC and we followed the following steps for all the initial conditions:
\begin{enumerate}
  \item Converting of the ".csv" file to ".dat" file.
  \item Remove any alphabets from the ".dat" file.
  \item Remove all irrelevant cycles retaining only the nontransient cycles of the third part (from both $V_{out}(t)$ and $V_{in}(t)$). The removed cycles, thus, include $R_{1-}$ cycles, the 1st, the 2nd part, the transient cycles ($\approx$ 21 cycles) of the 3rd part (of the $R_{123}$ cycles) and the $R_{3+}$ cycles. Hence, we are left with only 17979 nontransient cycles of the third part for analysis.
  \item We plot the hysteresis loops (i.e., $V_{out}(t)$ versus $I_{inp}(t)$, a constant factor of $V_{in}(t)$) and find the average loop.
  \item We find the area of the loop(s) for all the initial conditions of a particular noise strength.
  \item We find the average area for several noise strengths $T$ and finally plot the average (over all possible initial conditions) area $<\bar A>$  versus the noise strength (or in terms of their temperature $T$).
\end{enumerate}

\maketitle
\section{Experimental Results}

In our experiment, the output voltage $V_{out}(t)$ is analogous to the trajectory $x(t)$ of Eq. (2.2). The parameters of the plots, shown in Figs. 4-11, are all in dimensionless units. We perform the experiment in the earlier stated region of parameter space $(\gamma ,\tau)$ i.e., in the space of $(\{\frac{R_2C_1}{R_B}+C_3+\frac {R_2C_2}{R_A}\}(R_4/m)^{0.5},(R_4m)^{-0.5}f^{-1})$ and for $'F_0'$ or the amplitude of the drive current $I_0=(V_{in}^0R_4/V_0R_1)$ near about 0.2. 

\subsection{The two dynamical states}

Fig. 4 shows that, at $T=0$, for the same periodic drive $I_{inp}(t)$, we obtain two (and only two) possible kinds of trajectories depending upon the initial conditions taken. One of them oscillating with a large amplitude (LA state) and having a large phase lag with respect to the externally applied field $I_{inp}(t)$ and the other oscillating with a small amplitude (SA state) and having a small phase lag. In other words, for a given applied field $I_{inp}(t)$ at $T=0$, a fraction of initial conditions yield large amplitude output signals and the remaining small amplitude output signals. The fraction depends on the choice of the pair of parameters ($\gamma,\tau$). The plot of output signal with the input signal gives a hysteresis loop and in the present case, it gives two kinds of hysteresis loops, corresponding to two kinds of trajectories are obtained, as shown in Fig. 5. The mean area $\bar A =\overline{ \oint V_{out}(t)dI_{inp}(t)}$ of the hysteresis loops gives the energy absorbed by the system per period of the drive field (drive current).

\begin{figure}[htp]
\centering
\includegraphics[width=14cm,height=5cm]{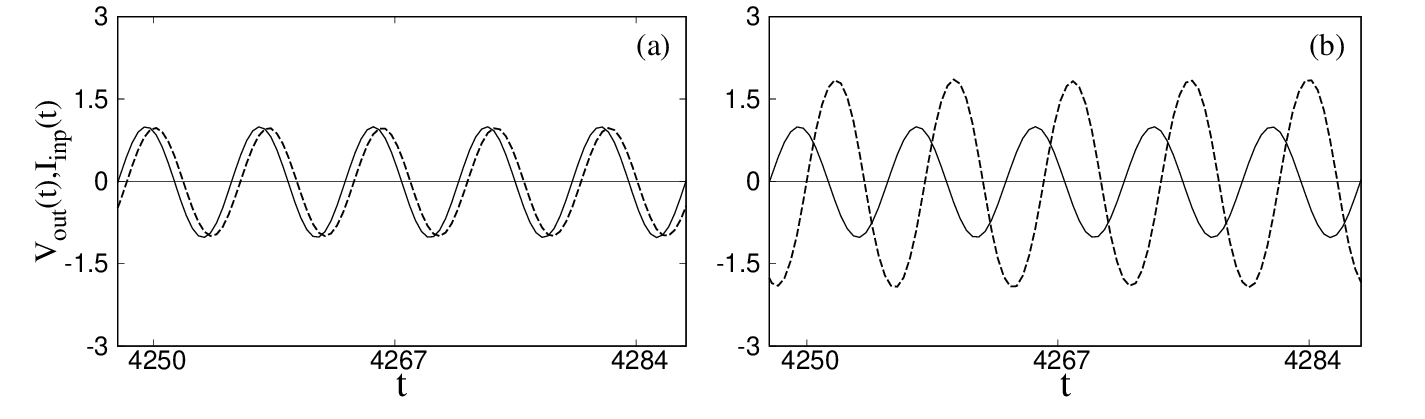}
\caption{Plot of the output signal (output voltage, shown by dotted line) and the input signal (input current, shown by thick line) in dimensionless units, where the trajectory of the output signal shows the (a) in-phase state when the initial condition $V_{out}(t=0)=+0.050$ and (b) out-of-phase state when the initial condition $V_{out}(t=0)=-1.81875$. Here $\gamma =0.1181$, $I_0=0.2$, $\tau=8.0$, $T=0$ and the amplitude of the input signal has been multiplied by a factor of 5.061. $t=4250$ in dimensionless unit corresponds to the time $t=0.100066$s when we have taken the parameter values given in Fig. 2.}
\label{fig4}
\end{figure}

\begin{figure}[htp]
\centering
\includegraphics[width=14cm,height=6cm]{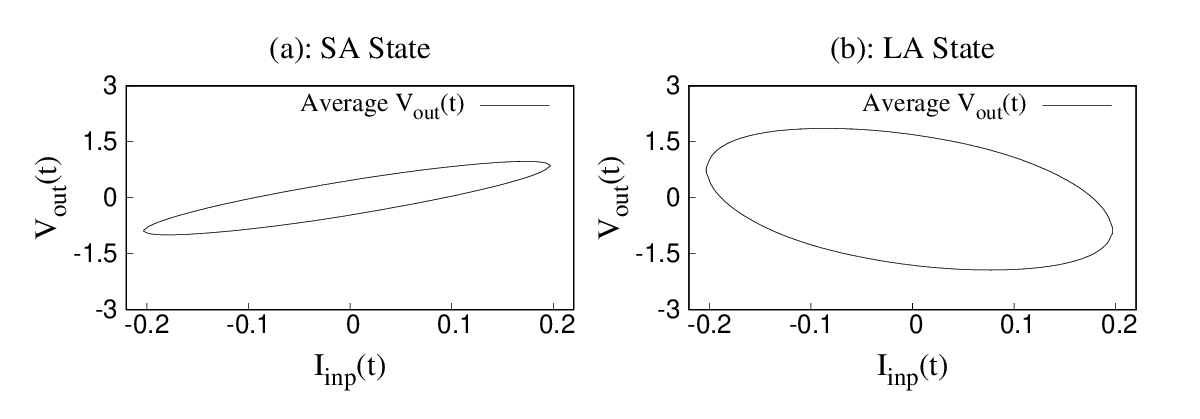}
\caption{Plot of the average hysteresis loop area for the (a) In-phase state and (b) Out-of-phase state.}
\label{fig5}
\end{figure}

Fig. 6 shows the regions of existence of the LA and SA states in the $(\tau-\gamma )$ plane. This figure is similar to the numerical result shown in Fig. 3 of Ref.\cite{wanda2015}. For any particular $\gamma$ value we see only the LA states in the small $\tau$ region. As $\tau$ is increased, SA states begin to appear at a fixed value of $\tau$. This particular $\tau$ value sets a boundary between the LA states-only region and the region of coexistence of the LA and SA states. This boundary is shown by a dotted line. As $\tau$ is further increased beyond a certain large value, we no longer see the LA states and we see only the SA states. This particular $\tau$ sets a boundary between the region of coexistence of the LA and SA states, and the SA states-only region. This boundary is shown by a thick line. For $\gamma > 0.1415$, it is difficult to take the readings since the output signal $V_{out}(t)$ fluctuates because the maximum amplitude of the output signal becomes around $3.14Vpp$. Around the maximum amplitude of $\pi/2$, the tracjectory of $V_{out}(t)$ broadens (is nearly a horizontal line) at the extreme points and the trajectory cannot decide whether to go to the channel 1 or channel 2,3  of the circuit shown in Fig. 3. However, if the maximum amplitude of $V_{out}(t)$ is slightly $>+\pi/2$ (or $<+\pi/2$) or slightly $<-\pi/2$ (or $>-\pi/2$), this problem of fluctuation disappears.

\begin{figure}[htp]
\centering
\includegraphics[width=9cm,height=6cm]{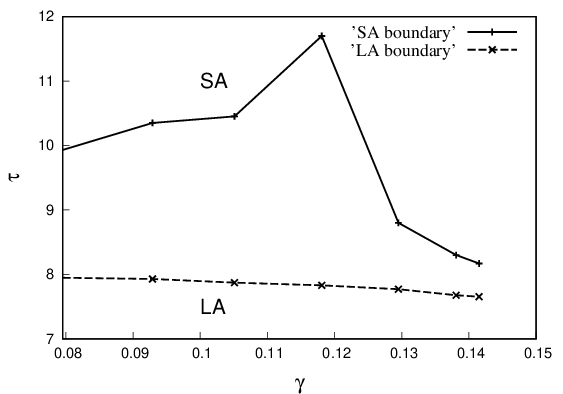}
\caption{Plot of the coexistence region of the LA and SA states in the $(\tau-\gamma )$ plane when the system is driven by a sinusoidal field of period $\tau$ and amplitude $I_0=0.2$ in a medium of uniform friction $\gamma $}
\label{fig6}
\end{figure}

\begin{figure}[htp]
\centering
\includegraphics[width=13cm,height=10cm]{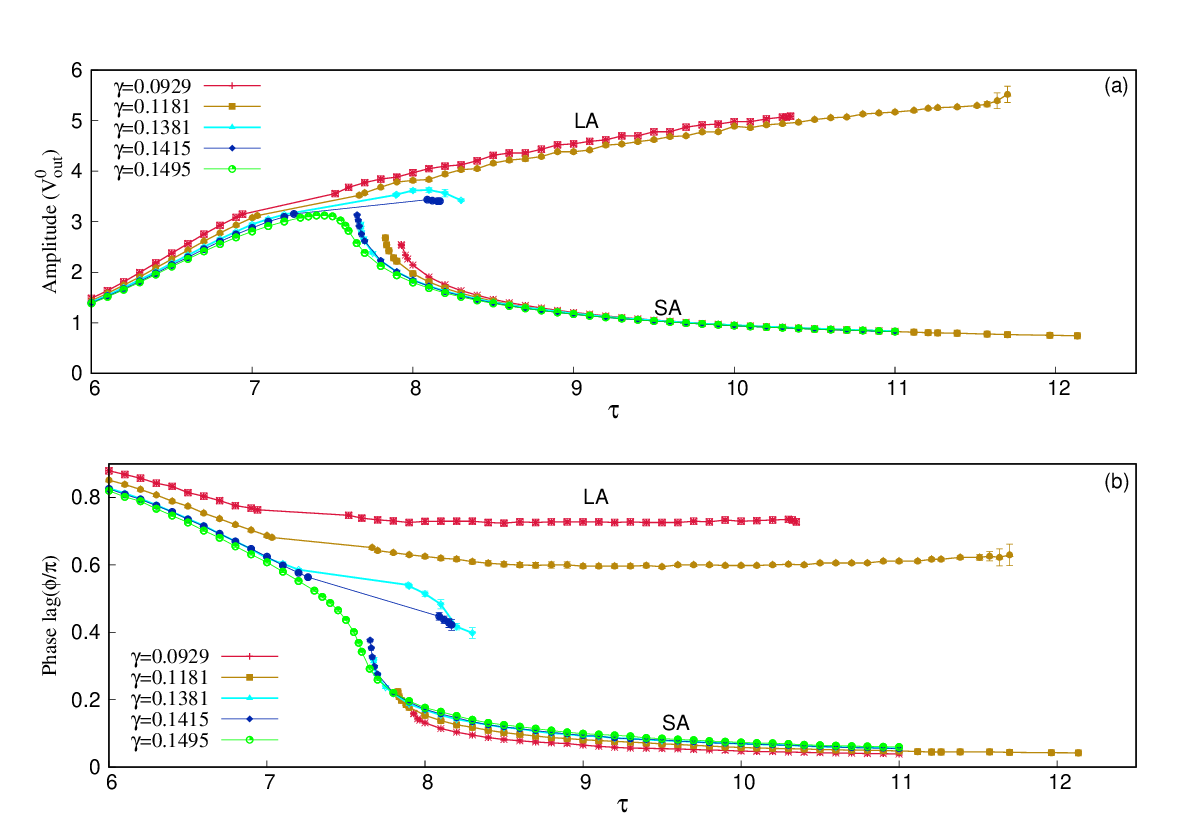}
\caption{Plot of the amplitudes $V^{0}_{out}$ (fig. (a)) and phase lags (fig. (b)) as a function of the period $\tau$ for different $\gamma $ values when $I_0=0.2$. In this plot, we can see two seperate branches for $\gamma =0.0795-0.1415$ and in this $\gamma $ values, we can see both the two states coexist. One corresponds to the LA states and the other to the SA states as shown in the plot. For $\gamma =0.1495$ we can see that both $V_{out}$ and $\phi $ change continuously as $\tau$ is varied.}
\label{fig7}
\end{figure}

Fig. 7 shows the peak-to-peak amplitude $V_{out}^0$ (Fig. 7a) and phase lag $\phi$ (Fig.7b) of the two dynamical states for different $\gamma $ values as a function of $\tau$. This figure is similar to the numerical result shown in Fig. 4 of Ref.\cite{wanda2015}. Here, the amplitude of the sinusoidal input current is 0.2. We find that both $V_{out}^0$ and $\phi $ change continuously as $\tau$ is varied for $\gamma $ values ranging from 0.0795 to 0.1415. However, there is a clear separation between the two LA  (upper) and SA (lower) branches in both the Figs. 7(a) and 7(b) for $0.0795\leq\gamma\leq 0.1415$. For $\gamma =0.1495$, we see only one kind of trajectory where the amplitude $V_{out}^0$ and phase lag $\phi$ change continuously as $\tau$ is varied for the entire range. Therefore, the distinction between the LA and the SA states disappears for $\gamma \ge 0.1495$.

\begin{figure}[htp]
\centering
\includegraphics[width=15cm,height=8cm]{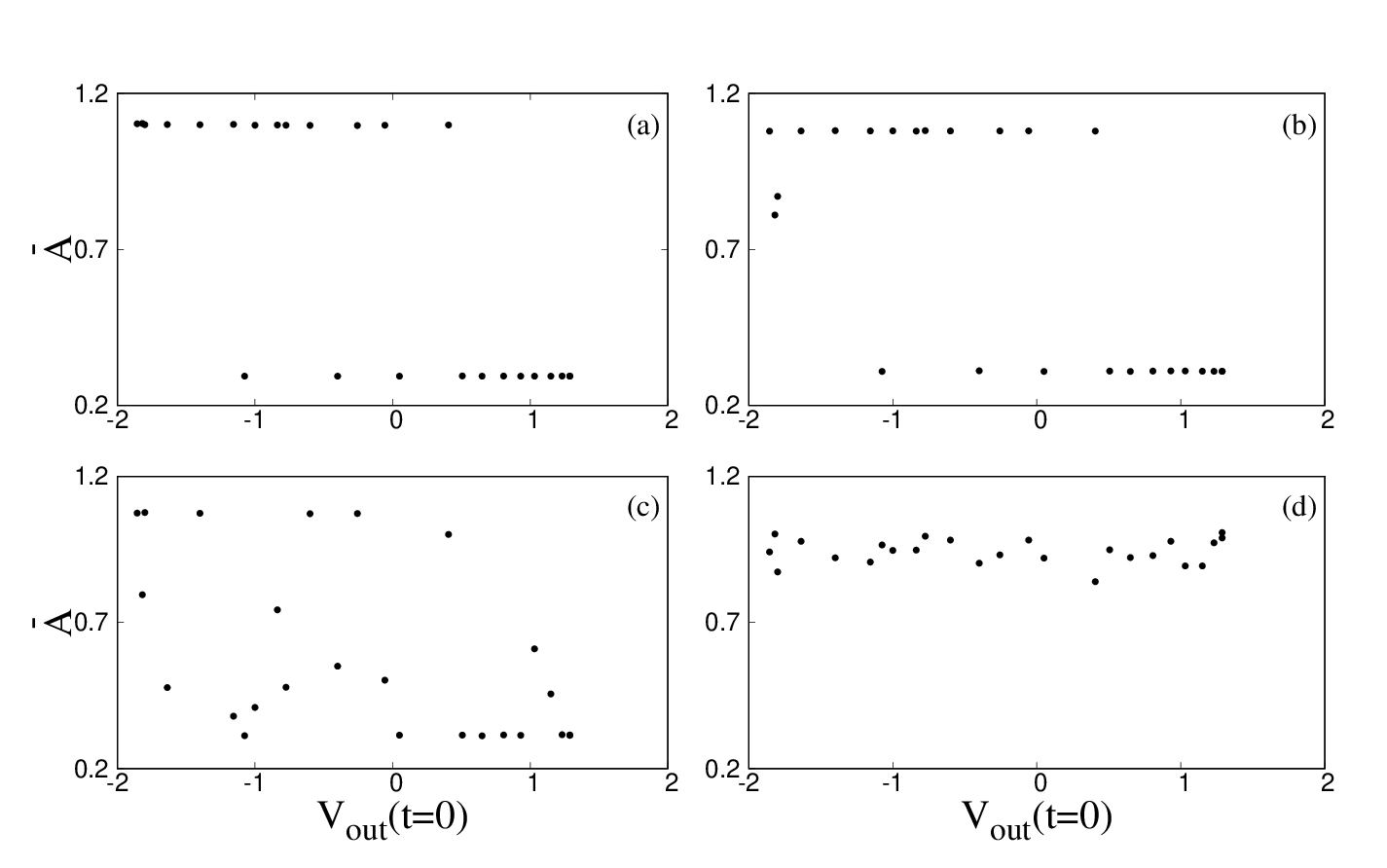}
\caption{Plot of $\bar A$ as a function of $V_{out}(t=0)$ for $T=0.0 $(a), $T=0.0595029$ (b), $T=0.080990$ (c) and $T=0.1999958$ (d) for $\tau= 8$, $I_0= 0.2$, and $\gamma = 0.1181$.}
\label{fig8}
\end{figure}

\begin{figure}[htp]
\centering
\includegraphics[width=15cm,height=8cm]{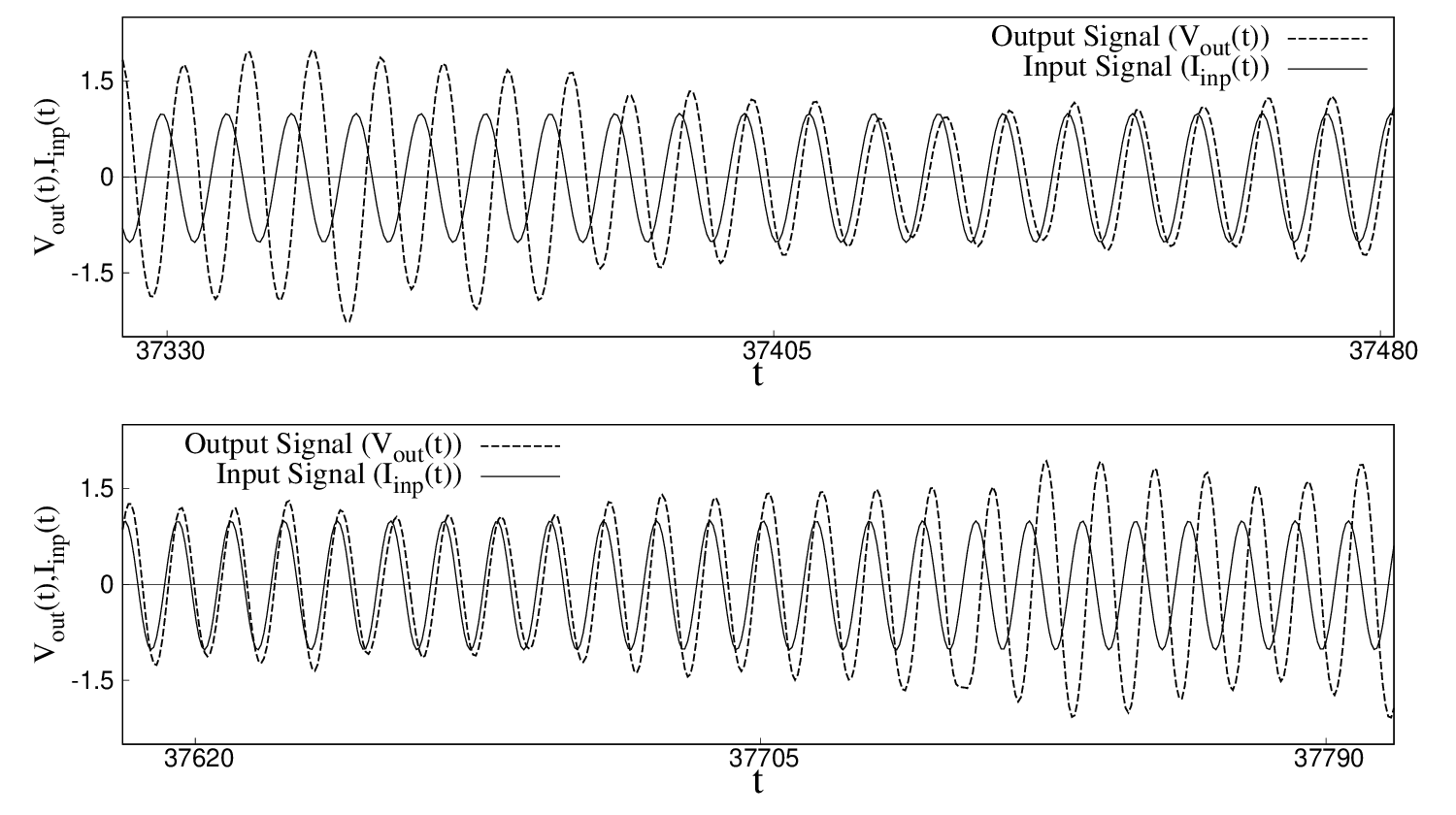}
\caption{Plot of the output and input signal in two parts (a) and (b) when T=0.1999958, $\tau= 8$, $I_0= 0.2$, and $\gamma = 0.1181$ and the amplitude of the input signal has been multiplied by a factor of 5.061. Fig. 9(a) shows the LA state jumps to the SA state and Fig. 9(b) shows the SA state jumps to the LA state.}
\label{fig9}
\end{figure}

Fig. 8(a) shows the plot between the average hysteresis loop area $(\bar A)$, average over 17979 number of cycles versus the initial condition $V_{out}(0)$ at $T=0$. From this plot, we see that the average hysteresis loop area is confined to two regions only i.e, one in which the $\bar A$=1.1 and the other in which $\bar A$=0.29. The $\bar A$=1.1 corresponds to the LA state with phase lag $\phi _1 \approx 0.63\pi$ and those with $\bar A$=0.29 corresponds to the SA state with phase lag $\phi _2 \approx 0.16\pi$. These two trajectories are very stable at low temperatures and they remain in the same state for any number of cycles taken. From Fig. 8(a), out of the total number of initial conditions that we have taken, nearly 48\% yield the SA state and remaining yield the LA state for $\tau=8.0$. For a given $\tau$ value, we have taken 25 initial conditions for each noise strength. Out of these 25 initial conditions, we obtain varying number of LA and the SA states for various $\tau$ values: For example, for $\tau=7.943$ we get 9 SA and 16 LA states, for $\tau=8.0$ 12 SA and 13 LA states, for $\tau=8.14$ 15 SA and 10 LA states and for $\tau=8.2$ 19 SA and 6 LA states.

Characteristically, depending on the values of ($\gamma,\tau$), at $T=0$, the two dynamical states of trajectories shown in Fig. 4, have their own basins of attraction but energetically one may be more stable than the other. Consequently, as additional provocations due to thermal fluctuations are included, transition between the two states takes place. Hence, the fraction of the large amplitude states, vis a vis the small amplitude states, changes.

\subsection{The stochastic resonance}

As the temperature is increased (by adding Gaussian white noise signal to the input voltage signal of Fig. 2 and in this experimental results, we have divide R by a factor 6.192 so that the temperature from the experimental results becomes comparable with that from the numerical results\cite {Saikia}), e.g., from $T=0$ to $T=0.0595029$ (Fig. 8(b)), the initial LA states begin to jump to the SA state, whereas the initial SA states remain in the same state. This shows that the SA state is more stable than the LA state for $\tau =8.0$, $\gamma =0.1181$ and $I_0=0.2$. The reason that we do not see a single band of pure SA states only is because the number of cycles (=17979) taken in our experiment is not enough to see the completion of transitions from all LA states to SA state. Note that in the numerical calculation of Ref.\cite{Saikia}, more than 200000 cycles of input signals were used to obtain their results.  As we increased the temperature (e.g., $T=0.080990$, Fig. 8(c)), not only the LA states jump to the SA state, SA states also begin to jump to the LA state. Fig. 9 shows the trajectory of the output and input signal where we can clearly see that LA jumps to the SA state (Fig. 9(a)) and SA jumps to the LA state (Fig. 9(b)). Thus the average hysteresis loop area (and hence $<\bar A>$, average over all possible initial conditions) begins its upward trend from a minimum. Fig. 8(d) shows a situation at $T=0.1999958$ when the average (over all possible initial conditions) hysteresis loop area, $<\bar A>=<\overline{ \oint V_{out}(t)dI_{inp}(t)}>$, becomes maximum.

\begin{figure}[htp]
\centering
\includegraphics[width=16cm,height=6cm]{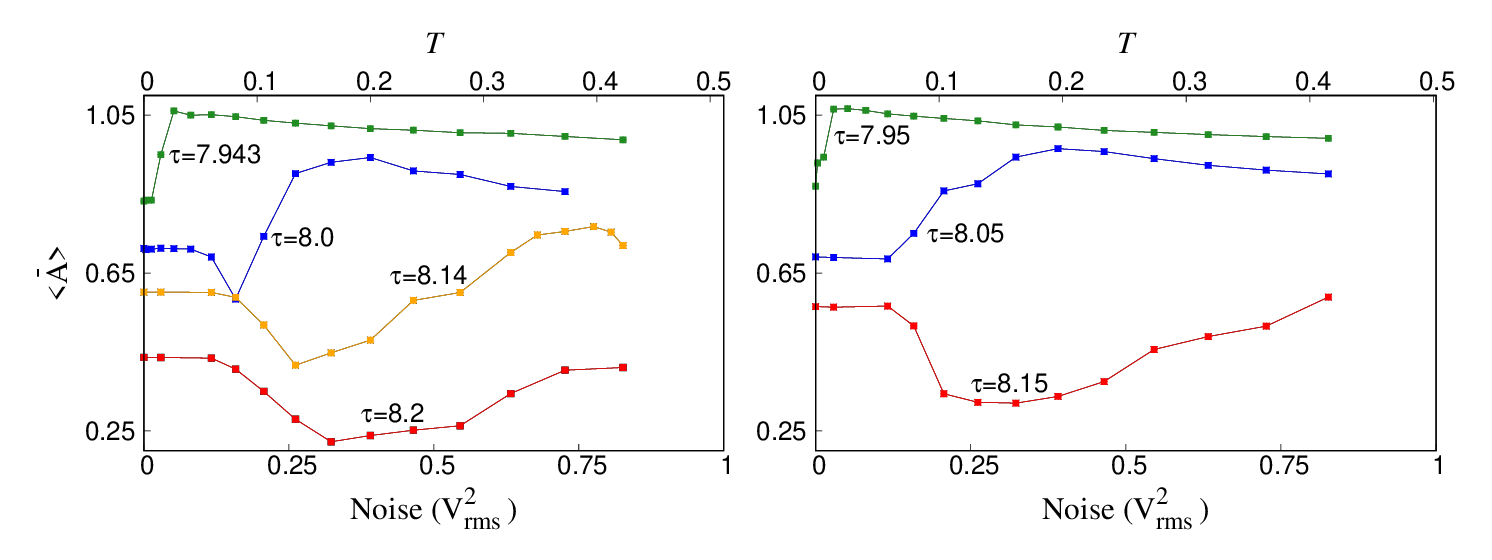}
\caption{Plot of $<\bar A>$ as a function of temperature V$_{rms}^2$ (and $T$) for various values of the period $\tau$ when $I_0= 0.2$, and (a) $\gamma = 0.1181$ (b) $\gamma = 0.1204$.}
\label{fig10}
\end{figure}

Figures 10(a) and 10(b) show how the average hysteresis loop area ($<\bar A> $, from the current-voltage characteristic of the circuit, Fig. 2) varies as a function of noise strength V$_{rms}^2$ (temperature $T$) for (a) $\gamma = 0.1181$ when $\tau=$ 7.943, 8.0, 8.14 and 8.2 and (b) $\gamma = 0.1204$ when $\tau=$ 7.95, 8.05 and 8.15 respectively.

For $\tau=$7.943, $<\bar A>$ peaks at very small $T$ as shown in Fig. 10(a). At such small temperatures, only intrawell transitions take place between SA and LA, and no interwell transitions were possible. Hence, peaking of $<\bar A>$ cannot actually be termed as SR in a periodic potential because the movement of the particle covers only a small part of a single well of the sinusoidal potential. It can be termed as SR in a single well potential. 

For $\tau=$8.0 (or $\tau=$8.14), Fig. 10(a) shows the nature of typical SR with interwell transitions around the peak of $<\bar A>$ and a characteristic initial dip. Even though we do not directly see the trajectory of the output signal oscillating in the next adjacent wells (due to interwell transitions), we infer its presence on the basis of the amplitudes of the output signal trajectory where the amplitude is more than $\pi$. When the amplitudes of the output signal trajectory is around $\pi$ with a reference dc offset voltage of $\approx 0$, we see that the trajectory jumps to a maximum $-ve$ rail voltage where the trajectory becomes a constant dc for few cycles of the $I_{inp}(t)$ and then it jumps back to the original reference dc offset voltage of $\approx 0$ and starts oscillating again. However, when it jumps to the maximum $+ve$ rail voltage, it stays there only where the trajectory becomes a constant dc. We save the data points, from the oscilloscope, for both, the intrawell and interwell transitions for moderate to the maximum limit of $T$, as shown in Fig. 10(a). For calculating $\bar A$, we analyse the signal and save only trajectories which oscillate with a reference dc offset voltage of $\approx 0$. These jumps in the dc voltage when the amplitudes of the output signal is around $\pi$, shows the presence of the interwell transition for $\tau\geq 8.0$. 

The presence of intrawell and interwell transtions along with the maximum $<\bar A>$ at an intermediate temperature  confirms the occurence of SR in sinusoidal potentials using the analog simulation experiment. For $\tau=$8.2, we see the initial dip of $<\bar A>$ and also rising of $<\bar A>$ as $T$ is increased further. However, there is a limitation in the circuit that we cannot increase the temperature more than $T=0.4231316$ since there are too many interwell transitions in the output signal and it is difficult to analyse the cycles required for calculating the $\bar A$.

Fig. 10(b) shows similar results as that obtained in Fig. 10(a) but with a different value of $\gamma $. $\tau=$ 7.95 in Fig. 10(b) is similar to $\tau=$ 7.943 from Fig. 10(a) where $<\bar A>$ peaks at very small $T$. Around the peak of $<\bar A>$, interwell transition is not possible since the output signal does not cover the full range of the single well in a sinusoidal potential and only intrawell transition is possible. Therefore the peaking of $<\bar A>$ can be termed as SR in a single well potential.

The period $\tau=$ 8.05 in Fig. 10(b) is similar to $\tau=$ 8.0 from Fig. 10(a) where $<\bar A>$ peaks at moderate $T$. For this $\tau$ value, intrawell and interwell transitions are present at moderate temperatures and intrawell transitions occur at a much smaller temperature compared to interwell transitions. As mentioned earlier, the presence of intrawell and interwell transtions along with the maximum $<\bar A>$ at an intermediate temperature  confirms the occurence of SR in sinusoidal potentials using the analog simulation experiment when $\tau=$ 8.05.
$\tau=$ 8.15 in Fig. 10(b) is similar to $\tau=$ 8.2 from Fig. 10(a) where $<\bar A>$ is still increasing and since we cannot increase the temperature further, therefore we cannot reach the point where $<\bar A>$ becomes maximum for this particular $\tau$ value.

\begin{figure}[htp]
\centering
\includegraphics[width=12cm,height=7cm]{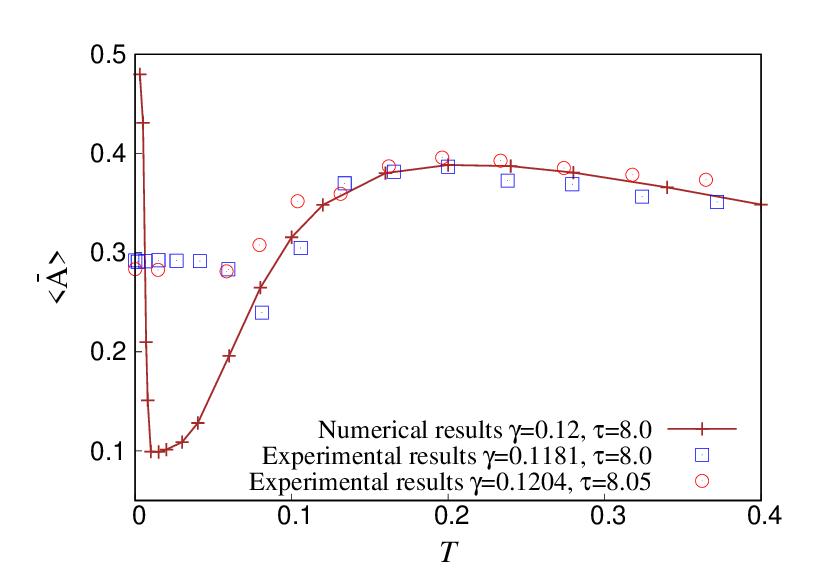}
\caption{Plot of $<\bar A>$ as a function of temperature $T$ where the numerical results ($\gamma = 0.12$, Fig. 8 from Ref. \cite {Saikia}) is compared with the (scaled) experimental results ($\gamma = 0.1181$ and $\gamma = 0.1204$).}
\label{fig11}
\end{figure}

Fig. 11 shows the plot between $<\bar A>$ as a function of $T$ where we have compared the numerical results, when $\gamma =0.12$ and $\tau =8.0$, with the experimental results, when $\gamma =0.1181$ and $\tau =8.0$, and when $\gamma =0.1204$ and $\tau =8.05$. From these three results, we see that the experimental results show a qualitative trend when compared with the numerical result. In this experimental results, we have multiplied $<\bar A>$ by a factor 0.41 so that $<\bar A>$ from the experimental results becomes comparable with that from the numerical results\cite {Saikia}.

\section{Discussion and Conclusion}

A sinusoidal potential system driven by an applied periodic current of small amplitude at a frequency close to its natural frequency of oscillation shows stochastic resonance. In this work, the average hysteresis loop area which represents the energy absorbed by the system per period of the drive field (drive current), is used as a quantifier of SR.

The experimental results are not exactly same as those of the numerical results but the qualitative trends are similar.  There are few difficulties in the experiment to obtain the solutions of the equation of motion. These shortcomings could have led to results slightly different from the numerical ones. Firstly, the initial conditions that we could access are not as complete as the ones taken in the numerical calculations. Therefore, the evaluation of $<\bar A>$ is statistically not as numerous as in the numerical experiment. This is more true for small temperatures where initial conditions determine for all time whether the states are SA or LA and hence correct averaging requires information about the fraction of the SA and LA states. However, at higher temperatures, due to frequent transitions between SA and LA states the memory of initial conditions is lost and thus the evaluation of $<\bar A>$ is not affected. Also, the other initial condition $\frac{d}{dt}V_{out}(t=0)$ is left arbitrary. Secondly, we cannot increase the temperature to larger values as too many interwell transitions occur at such noise strengths. Oscillations in potential wells other than the initial one lead to nonzero finite offset voltages, sometimes comparable to the power supply voltages to the ICs. Also, since the sine converter cannot convert any $\theta $ values beyond $\pm \pi$, due to interwell transitions, the trajectories jump to the maximum rail voltages.

In summary, using the analog circuit, SR is, indeed, found to occur in sinusoidal potentials. There are many difficulties we face in order to achieve the desired results. Some of the difficulties has been mentioned in Section III (b), (c) and (d). Calculating the mean hysteresis loop area is a tedious procedure. Here, in a single run, we obtain a mixture of hysteresis loops corresponding to several large amplitude and small amplitude trajectories. In order to obtain a mean loop, one needs to adopt appropriate but ad-hoc criteria. Of course, there is no foolproof criterion and one has to live with an error. However, this error is not large enough to change the qualitative trend of the curves in Fig. 10(a) (or Fig. 10(b)), a result of several months of real-time effort. We have taken enough precautions in obtaining the raw data and its analysis. Even though the results may not be exactly reproducible quantitatively, due to variation of the characteristics of the electronic components with time, the qualitative trend of the curves of Fig. 10(a) (or Fig. 10(b)) is definite. That is, the occurrence of stochastic resonance in sinusoidal potentials is, undoubtedly, verified to be correct.

\end{document}